\documentclass[twoside,twocolumn]{article}

\usepackage{blindtext}
\usepackage{siunitx}
\usepackage{graphicx}
\usepackage{tabularx}
\usepackage{bm,amsmath,amsfonts,amssymb,amsthm}
\usepackage[sc]{mathpazo}
\usepackage[square,sort&compress,comma,numbers]{natbib}
\usepackage{microtype} 

\usepackage[english]{babel} 

\usepackage[hmarginratio=1:1,top=20mm,columnsep=20pt,margin=15mm]{geometry} 
\usepackage[small,labelfont=bf,up]{caption} 
\usepackage{booktabs} 

\usepackage{lettrine} 

\usepackage{enumitem} 
\setlist[itemize]{noitemsep} 

\usepackage{abstract} 

\usepackage{titlesec} 
\renewcommand\thesection{\Roman{section}} 
\renewcommand\thesubsection{\roman{subsection}} 
\titleformat{\section}[block]{\large\scshape\centering}{\thesection.}{1em}{} 
\titleformat{\subsection}[block]{\large}{\thesubsection.}{1em}{} 

\usepackage{titling} 

\usepackage{hyperref} 

\title{Universal motion of mirror-symmetric microparticles in confined Stokes flow} 
\author{%
\textsc{Rumen N. Georgiev and Sara O. Toscano}\\ 
\normalsize Process and Energy Department, Delft University of Technology, The Netherlands \\ \\
\textsc{William E. Uspal}\\
\normalsize Department of Mechanical Engineering, University of Hawai\'i at Manoa, USA \\ \\
\textsc{Bram Bet, Sela Samin and Ren\'e van Roij}\\
\normalsize Institute for Theoretical Physics, Utrecht University, The Netherlands \\ \\
\textsc{Huseyin Burak Eral}\thanks{Corresponding author: H.B.Eral@tudelft.nl}\\
\normalsize Process and Energy Department, Delft University of Technology, The Netherlands\\
\normalsize and\\
\normalsize Van't Hoff Laboratory for Physical and Colloid Chemistry, Utrecht University, The Netherlands \\ \\  
}

\renewcommand{\maketitlehookd}{%
\begin{abstract}
\noindent Comprehensive understanding of particle motion in microfluidic devices is essential to unlock novel technologies for shape-based separation and sorting of microparticles like microplastics, cells and crystal polymorphs. Such particles interact hydrodynamically with confining surfaces, thus altering their trajectories. These hydrodynamic interactions are shape-dependent and can be tuned to guide a particle along a specific path. We produce strongly confined particles with various shapes in a shallow microfluidic channel via stop flow lithography. Regardless of their exact shape, particles with a single mirror plane have identical modes of motion: in-plane rotation and cross-stream translation along a bell-shaped path. Each mode has a characteristic time, determined by particle geometry. Furthermore, each particle trajectory can be scaled by its respective characteristic times onto two master curves. We propose minimalistic relations linking these timescales to particle shape. Together these master curves yield a trajectory universal to particles with a single mirror plane.
\end{abstract}
}

\begin{document}

\maketitle


\lettrine[nindent=0em,lines=3]{S}{}eparation on the microscale is a persistent industrial challenge: pharmaceutical crystal polymorphs \cite{Bauer2001,Shet2004}, specific strains of yeast cells in the food industry \cite{Piel2009}, mammalian cells \cite{Ginzberg2015} and microplastic pollutants \cite{Thompson2005, Taylor2016} all come in different shapes, yet comparable sizes. Advances in microfluidics have resulted in robust and high throughput methods for micron-scale segregation. These techniques rely on external force fields \cite{Ding2014, Lenshof2010}, sorting based on fluorescence \cite{Mage2019}, intricate separator geometries \cite{Nivedita2013, Son2017, Kim2016, Jiang2019, Behdani2018, Li2017, Russom2009, Mach2011, Huang2004} or carriers with non-Newtonian behaviour \cite{Raoufi2019}. An alternative approach towards microscale separation is to leverage the long-range hydrodynamic interactions emerging from fluid-structure coupling \cite{Hur2011, masaeli2012continuous}. By tuning these interactions particle trajectory can be controlled, thus enabling separation \cite{Uspal2013}.

A model system common in microfluidic applications, exhibiting such interactions, is confined Stokes flow in a Hele-Shaw cell. In it, particles or droplets are sandwiched between a pair of confining walls of a shallow microfluidic channel and are subjected to creeping flow  \cite{C7CS00374A}. Owing to the shallowness of the cell, the flow is effectively two-dimensional \cite{batchelor_2000}. What is more, the particle scatters the surrounding fluid, creating a dipolar flow disturbance, which decays with $1/r^2$, where $r$ is the distance from the particle centre. This flow disturbance strongly couples the particle to its surroundings. Experimentally, creating and driving particles in shallow channels has become widely accessible with the advent of microfluidics and soft lithography \cite{Teh2008, Zhu2017, Shang2017, Dendukuri2006, Dendukuri2007, Dendukuri2009, Chung2007}. Their easy fabrication and versatile out-of-equilibrium behaviour make particles in confined Stokes flow an interesting toy system for the study of flow-mediated separation and self-assembly \cite{Ge2019,Uspal2014}.  

Utilizing long-ranged hydrodynamic interactions (HIs), Beatus \textit{et al.} demonstrated how trains of `pancake' droplets flow along a Hele-Shaw cell as out-of-equilibrium 1D crystals \cite{Beatus2006,beatus2012physics}. In a similar experiment, Shen and co-workers compare the dynamics of clusters comprising 2 or 3 droplets as they interact near or far away from the side walls of the cell \cite{shen2014dynamics}. The presence of a side wall breaks the symmetry of the system and induces transversal motion of the cluster. Cross-streamline migration is also present if the symmetry of an individual particle, rather than that of an ensemble of particles, is reduced. A particle with two planes of mirror symmetry, such as a rod \cite{Berthet2013,Nagel2018} or a symmetric disk dimer \cite{Uspal2013}, also moves towards one of the side walls of a Hele-Shaw cell, provided its long axis is neither normal, nor parallel to the flow. As one such particle approaches the channel boundary, it begins to interact with its hydrodynamic image \cite{uspal2012scattering}, the flow symmetry is reduced even further and the particle begins to rotate. All three modes of motion, namely, rotation, streamwise and cross-streamwise translation, are also present when an asymmetric disk dimer is far away from any side walls as demonstrated by Uspal, Eral \& Doyle \cite{Uspal2013}. Evidently, screened hydrodynamic interactions give rise to non-trivial behaviour not only in particle ensembles \cite{Schneider2011,Green2018,Cui2002,Schiller2015,shani2014long}, but also in single-particle systems with broken symmetry {\cite{DuRoure2019,Fiorucci2019,Chakrabarty2013, Bechert2019,Gruziel2018,Cappello2019,Sowicka2013}. A first step towards the development of low-cost flow separators requires understanding the relation between the geometry of one such particle and its trajectory in confined Stokes flow. 
	
	In this study, we combine theoretical and experimental approaches to investigate how particle shape can be tailored to induce self-steering under flow in quasi-2D microchannels. Controlling the motion of a particle in flow facilitates its separation. To this end, we use optical microscopy to track the in-plane motion of a variety of particles with a single mirror plane subjected to creeping flow in a shallow microfluidic channel. The mirror plane is perpendicular to the top and bottom walls of the channel and bisects the particle in two identical pieces (white dashes in Fig. \ref{fig:1} (a)-(d)). Through finite element calculations we link the shape-dependent dynamics of the particles to the flow disturbances they create as they lag the far-field flow. Using Stokes linearity and the force-free nature of the particles, we collapse their re-orientation and cross-streamwise dynamics onto two master curves. We accomplish this collapse by scaling each particle's angular and transversal velocities by two characteristic times. Finally, through minimalistic scaling relations we link these timescales to a particle's geometrical parameters including, but not limited to, area, moment of inertia and length. Our scaling arguments predict the characteristic times from both experiments and finite element computations up to a factor on the order of unity. This good agreement among experiments, simulations and scaling arguments is a strong indication that the observed dynamics is universal to mirror-symmetric particles in quasi-2D Stokes flow.
	
	\begin{figure}[ht!]
		\centering
		\includegraphics[width=\linewidth]{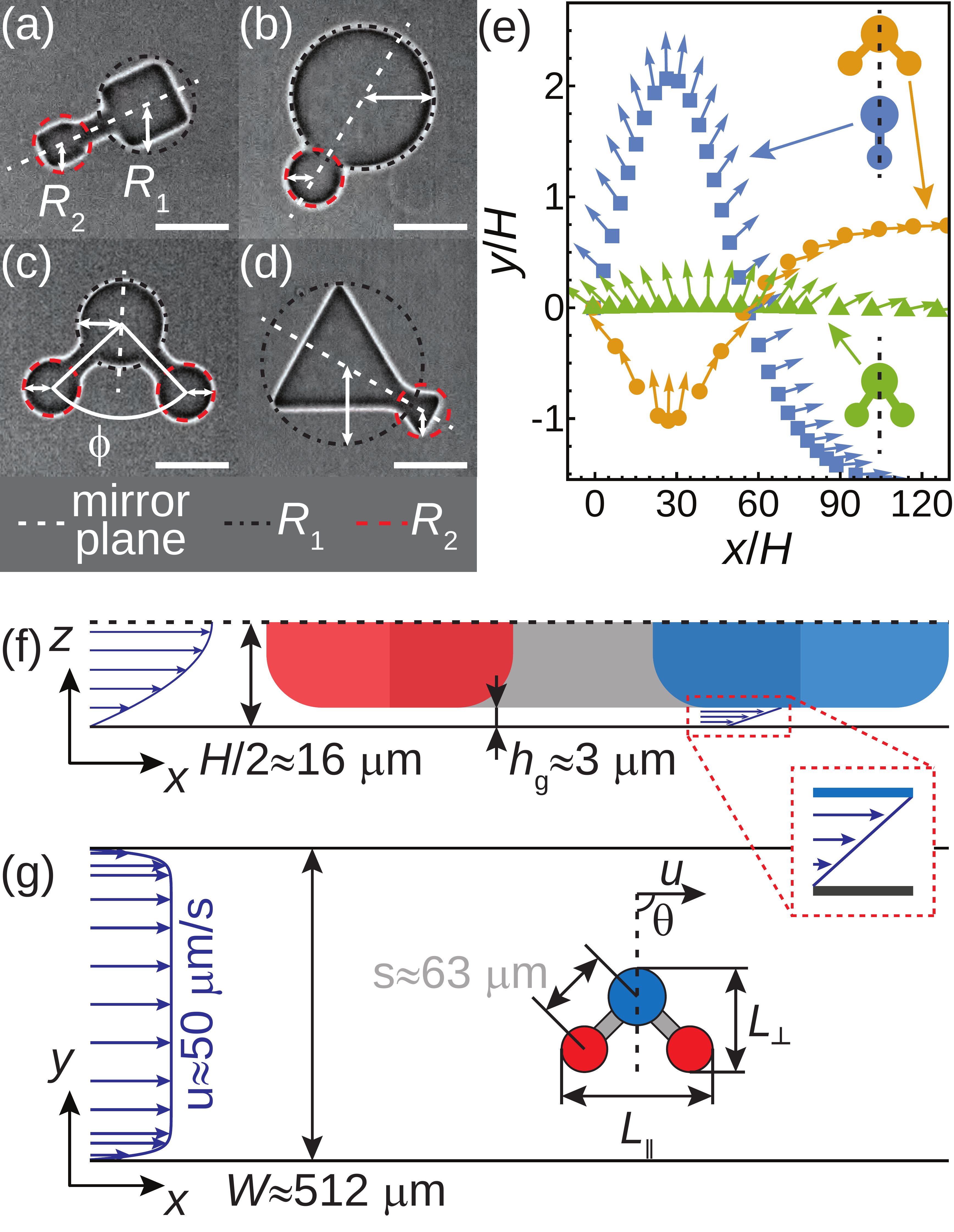}
		\caption{\textbf{Mirror-symmetric particles in quasi-2D Stokes flow.} Stop-flow lithography \cite{Dendukuri2007} produces strongly confined microparticles with various shapes in a Hele-Shaw cell (a-d). We investigate particles with a single mirror plane, each consisting of two or three simple building blocks such as disks, squares or triangles, connected with rigid shafts. These particles are a useful toy system to study how the geometry of a particle determines its trajectory. We demonstrate this strong shape dependence by comparing the trajectories of three particles with $R_1/R_2=1.5$: from top to bottom a trimer with $\phi=\SI{90}{\degree}$, a dimer and a trimer with $\phi=\SI{68}{\degree}$ (e). The small arrows denote the orientation of the particles. The trajectories are obtained via 3D finite element calculations. We assume a planar Poiseuille profile along the height of the channel and Couette flow in the thin lubrication gaps with height $h_\textrm{g}$ (f). Due to channel symmetry, we only present half of a Hele-Shaw cell with particle to scale. Upon depth-averaging, we arrive at the so-called Brinkman flow with steep velocity gradients near the horizontal walls and constant velocity $u$ along most of the channel width (g). In this top view the particle is magnified 2.5 times. The streamlines in all three flow profiles are represented by horizontal blue arrows. Scale bars are \SI{50}{\micro\metre}.}
		\label{fig:1}
	\end{figure}
	
	\begin{figure*}[ht!]
		\includegraphics[width=\linewidth]{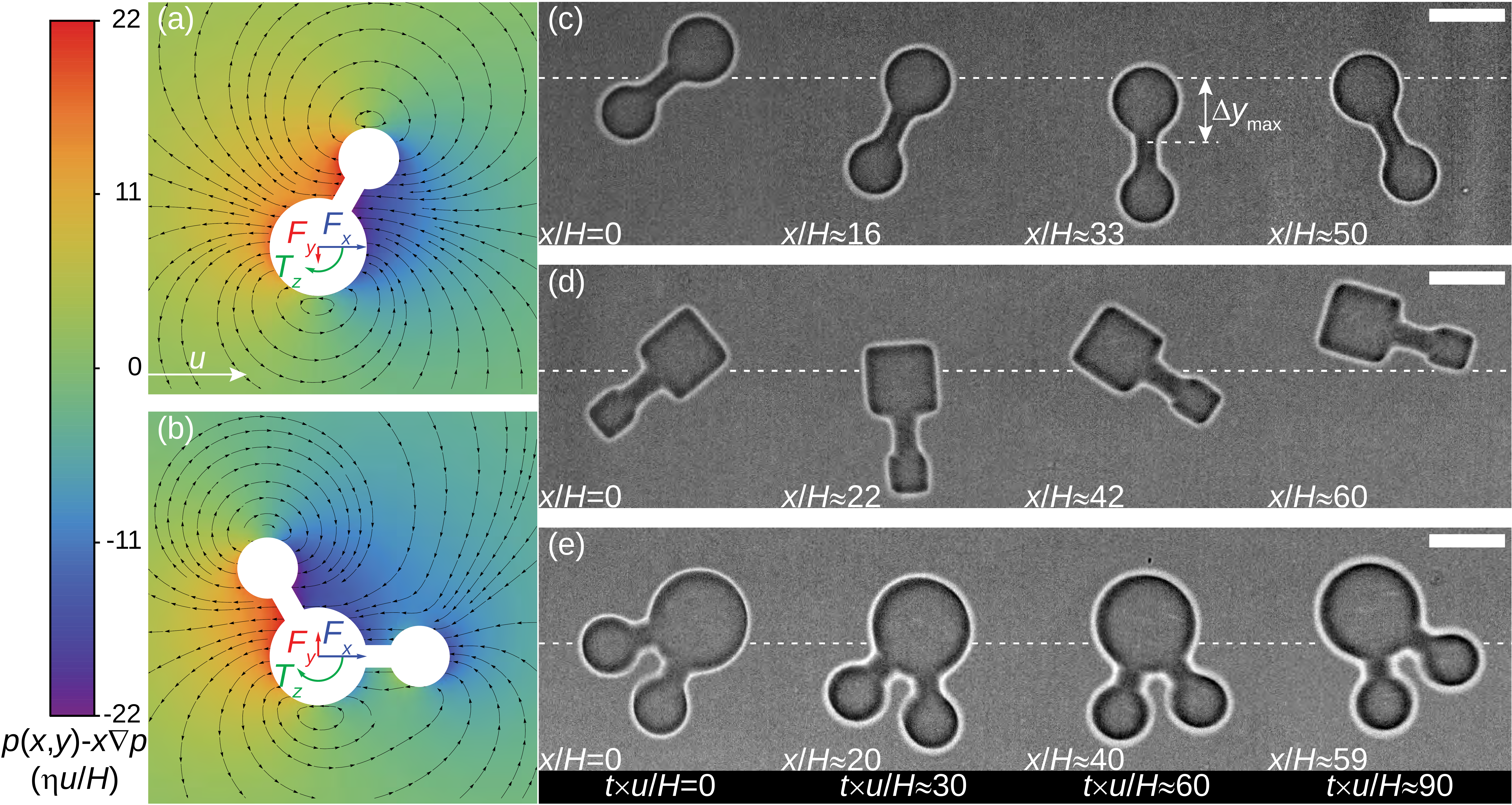}
		\caption{\textbf{Particle-induced flow disturbances in a Hele-Shaw cell.} As the particle thickness $H_\mathrm{p}=H-2h_\mathrm{g}$ is comparable to the channel height $H_\mathrm{p}/H\simeq0.8$, the particle lags the surrounding flow, creating shape-specific velocity and pressure disturbances (cf. arrows and density plots in a and b). As the disturbances differ, so too do the hydrodynamic forces and torque acting on each particle differ. While the streamwise forces $F_x$ on a dimer and a trimer have similar magnitudes  (horizontal blue arrows), the drift forces $F_y$ and torques $T_z$ acting on them differ (vertical red arrows and clockwise green arcs, respectively).  This shape-dependence of the forces and torque results in distinct linear and angular velocities, which manifest themselves in the different trajectories followed by different particles (cf. c, d and e). The orientation and scaled position $x/H$ as function of scaled time $t\times u/H$  are strongly dependent on particle shape. The disturbances to the pressure and velocity fields, as well as the forces and torques on the particles, are calculated using a 3D finite element scheme \cite{Bet2018}. In all sub-figures the flow is from left to right as denoted by the white arrow in a. Scale bars are \SI{50}{\micro\metre}.} \label{fig:2}
	\end{figure*}
	
	To produce strongly confined polymeric particles with distinct shapes in a Hele-Shaw cell we use stop-flow lithography (SFL) \cite{Dendukuri2007}, as depicted in Fig. \ref{fig:1} (a)-(d). In a nutshell, SFL creates particles by projecting the image of a mask onto a photoreactive fluid. We choose dimeric and trimeric particles, composed of, respectively, two or three simple shapes connected by rigid shafts. The building blocks for dimers are either disks, triangles or squares (Fig. \ref{fig:1} (b), (d) and (a)), while those for trimers are always disks (Fig. \ref{fig:1} (c)). In both cases one of the building blocks is larger with a size ratio $\kappa\equiv R_1/R_2$, where $1<\kappa\leq3$ and $R_2$ is the radius of the circle escribing the smaller shape. This asymmetry in the particle ensures its rotation even far away from any side walls \cite{Bretherton1962}. The trimers have an additional geometrical parameter, namely, the angle $\phi$ formed between the three disks (Fig. \ref{fig:1} (c)). The vertex of $\phi$ is defined as the centre of the larger disk, while the two rays starting from it point to the centres of the smaller equally-sized disks. By changing $\phi$ we gain additional control over the dynamics of the particles (Fig. \ref{fig:1} (e)). The geometry of the particle profoundly influences its trajectory: particles with identical starting positions, yet slightly different geometries, follow dramatically different paths, as demonstrated numerically in Fig. \ref{fig:1} (e).
	
	As the particles are created \textit{in situ}, we directly track their motion in the viscous fluid by moving the stage of an optical microscope. We set the system in motion by applying a small pressure drop across the channel, thus inducing creeping flow with a Reynolds number $Re\sim10^{-5}$. This flow regime, together with the large aspect ratio of the channel $W/H>15$, allows us to average out the parabolic profile expected along the channel height (Fig. \ref{fig:1} (f)). Thus, the particle is effectively subjected to an in-plane potential flow with steep velocity gradients near the side walls of the channel and a constant velocity $u$ for most of its width \cite{Bruus2011} (Fig. \ref{fig:1} (g)).
	
	Apart from preventing sticking, the fluid layers with thickness $h_\mathrm{g}$ present above and below the particle strongly affect its motion (Fig. \ref{fig:1} (f) and its inset). As the particle moves along the channel with a longitudinal velocity $\dot{x}$, it experiences additional drag, because it shears the lubricating fluid in the gaps. Due to the strong particle confinement the velocity profile in the gaps is close to linear \cite{Nagel2018,Bet2018}, allowing us to assume Couette flow in the gaps (Fig. \ref{fig:1} (f)). The drag from the confining walls $F_{x,\mathrm{w}}$ scales with $2\dot{x}\eta/h_\mathrm{g}$ and slows down the particle, where $\eta$ is the dynamic viscosity of the fluid. Furthermore, it ensures the particle is confined to the plane of the flow, because any tilt or out-of-plane motion results in additional force acting on either face of the particle. Thus, the particle exhibits three degrees of freedom: translation along the length $x$ and width $y$ of the channel and in-plane rotation $\theta$ (Fig. \ref{fig:1} (g)). 
	
	The particle lags the flow, perturbing the velocity field, and as a result pressure builds up on the upstream particle surface. This flow disturbance is strongly dependent on the particle shape (cf. (a) and (b) in Fig. \ref{fig:2}). To illustrate this phenomenon, we use finite element computations \cite{Bet2018} to calculate the forces and torque acting on two distinctly shaped particles with $\kappa=1.6$: a dimer and a trimer with $\phi=\SI{120}{\degree}$. We impose a unidirectional inlet flow with height-averaged velocity $u$ and prescribe a longitudinal velocity $\dot{x}=u/2$ to each particle. We orient the particles in such a way that their mirror axes form an angle $\theta=\SI{60}{\degree}$ with the flow. The particle heights $H_\mathrm{p}$ in both cases are equal and comparable to the channel height $H$, with $H_\mathrm{p}/H\sim0.8$. While the longitudinal forces $F_x$ acting on the two shapes are identical ($F^\mathrm{D}_x/F^\mathrm{T}_x=0.99$), the torques differ \--- the dimer experiences a smaller torque $T^\mathrm{D}_x/T^\mathrm{T}_x=0.81$. The superscripts `D' and `T' refer to `dimer' and `trimer'. The difference in the transversal forces $F_y$ is even more evident, as its direction also changes: $F^\mathrm{D}_y/F^\mathrm{T}_y=-0.67$. This disparity can be traced back to the pressure disturbance created by each particle \--- the larger the disturbance, the larger the forces.
	
	\begin{figure*}[ht!]
		\includegraphics[width=\linewidth]{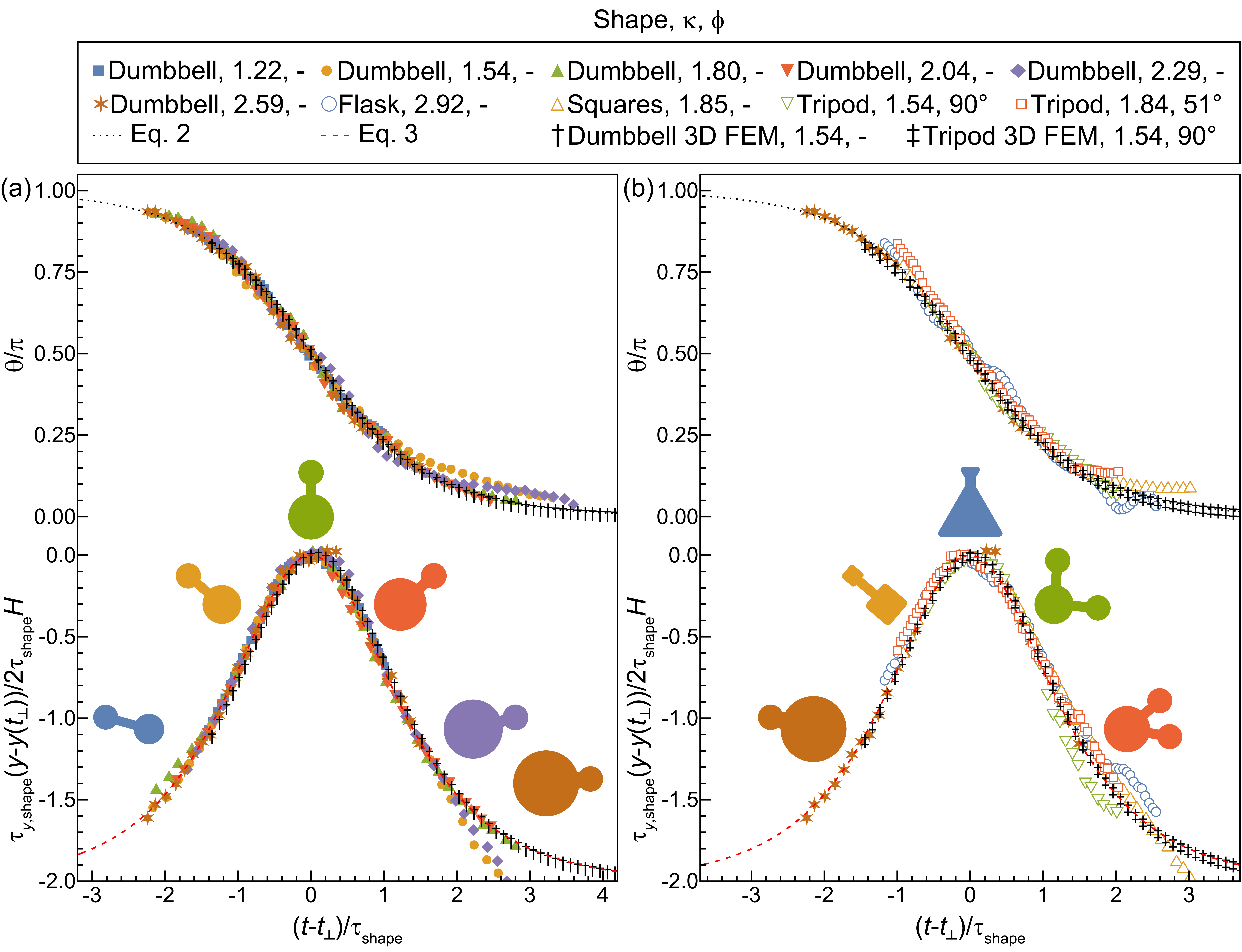} 
		\caption{\textbf{Universal behaviour of mirror-symmetric particles.} Regardless of their detailed shape, all studied particles follow a universal trajectory. They exhibit the same quantitative behaviour as long as we take into account two characteristic times, $\tau$ and $\tau_y$, scaling their modes of motion \cite{Bet2018}: exponentially-decaying rotation $\theta(t)$ to orient with the big disk upstream (top curves) and bell-shaped translation in the lateral direction $y\left(t\right)-y\left(t_\perp\right)$ (bottom curves). The only geometrical element common to all studied particles is their single plane of mirror symmetry. In all cases, the error bars denoting experimental uncertainty are smaller than the symbols and are omitted. The particles and their motion is sketched in the middle section of the figure. In the legend disk, square and triangle dimers are dubbed `dumbbell', `squares' and `flask' for brevity. Disk trimers are denoted as `tripod'.} \label{fig:3}
	\end{figure*}
	
	The shape-dependence of the disturbances manifests itself in the distinct dynamics of different particles, as shown in Fig. \ref{fig:1} (e). To demonstrate this distinction experimentally, we compare the motion of three particles with different shapes, which have one and the same initial position and orientation, $x/H$, $y/H$ and $\theta_0=7\pi/9$, respectively (Fig. \ref{fig:2} (c), (d) and (e)). While all three particles rotate to orient their larger building block upstream, only the dimers experience a significant lateral drift. Nagel \textit{et al.} \cite{Nagel2018} report a similar coupling between longitudinal and transversal motion for symmetric rods, which drift at a constant velocity as they flow downstream. However, cross-streamwise motion is orientation-dependent, resulting in a non-linear cross-stream trajectory when an asymmetric dimer rotates: as our particles become perpendicular to the flow, their transversal velocities diminish. Moreover, after acquiring this perpendicular orientation both particles change the direction of their lateral motion (cf. panel 3 in Fig. \ref{fig:2} (c) and panel 2 in Fig. \ref{fig:2} (d)). The coupling between rotation and translation explains why the disk dimer moves further away from its initial position $\Delta y_\mathrm{max}(t\times u/H=60)\sim 1.5H$ compared to the square dimer, which covers half of that distance in half the time (cf. panel 3 in Fig. \ref{fig:2} (c) and panel 2 in Fig. \ref{fig:2} (d)). Due to its slower rotation, the disk dimer spends a longer time crossing streamlines before orienting perpendicular to the flow and starting to move in the opposite direction. This reasoning does not, however, answer the question why the trimer experiences negligible drift, even though its rotational velocity is comparable to that of the disk dimer.
	
	Evidently, the observed coupling among the modes of translation and the rotation is hallmark of low-symmetry particles in a flow \cite{Russel1977}. Mathematically, we represent this inter-dependence using a resistance tensor $\textsf{\textit{R}}_\mathrm{p}$, a symmetric matrix with size equal to the number of degrees of freedom a particle exhibits (Supplementary Text 1A). The resistance tensor relates the hydrodynamic forces and torque a stationary fluid $u=0$ exerts on a particle, which translates through it with velocities $\dot{x}$ and $\dot{y}$, while also rotating at a rotational velocity $\dot{\theta}$ \cite{brenner1963stokes, brenner1964stokes}:
	\begin{equation}\label{eq:1}
	\begin{pmatrix}
	F_x\\
	F_y\\
	T_z
	\end{pmatrix}=-\eta\textsf{\textit{R}}_\mathrm{p}
	\cdot
	\begin{pmatrix}
	\dot{x}\\
	\dot{y}\\
	\dot{\theta}
	\end{pmatrix}, \text{with}\ \textsf{\textit{R}}_\mathrm{p}\sim
	\begin{pmatrix}
	l_{xx}&l_{xy}&l_{x\theta}^2\\
	l_{yx}&l_{yy}&l_{y\theta}^2\\
	l_{\theta x}^2&l_{\theta y}^2&l_{\theta\theta}^3
	\end{pmatrix}.
	\end{equation}
	We present each component of $\textsf{\textit{R}}_\mathrm{p}$ in terms of arbitrary length scales $l_{ij}$ to demonstrate one of its defining features \--- much like Stokes flow itself, the resistance tensor is time-independent and defined purely by geometry. If the particle possesses only a single mirror plane, all nine components of $\textsf{\textit{R}}_\mathrm{p}$ are generally non-zero, reflecting the entwined nature of its modes of motion (Supplementary Text 1C). Conversely, for a rod the $l_{ij}^2$ components become zero, since its coupled translational modes are unaffected by rotation. Particles with an even higher symmetry such as disks have all three modes independent of each other and their resistance tensors are diagonal matrices.
	
	Utilizing the concept of the resistance tensor together with Stokes linearity, we recently derived equations of motion for a force-free mirror-symmetric particle subjected to confined Stokes flow \cite{Samin2018} (Supplementary Text 1B). Both equations, as presented in \cite{Samin2018}, seemingly depend on the initial orientation of the particle $\theta_0$. However, once we realize Stokes flow is time-reversible, $\theta_0$ becomes an arbitrary reference angle. For convenience, we set $\theta_0=\pi/2$, resulting in: 
	\begin{equation}\label{eq:2}
	\theta\left(t\right)=2\arctan\left[\exp\left(-\frac{t-t_\perp}{\tau}\right)\right]
	\end{equation}
	and
	\begin{equation}\label{eq:3}
	y\left(t\right)=y\left(t_\perp\right)+2H\frac{\tau}{\tau_y}\left[\mathrm{sech}\left(\frac{t-t_\perp}{\tau}\right)-1\right],
	\end{equation}
	where $t_\perp=t\left(\theta=\pi/2\right)$ denotes the time at which the particle is perpendicular to the flow. The two timescales, $\tau$ and $\tau_y$, are characteristic for the re-orientation and cross-stream migration of each particle. Numerically, they can be computed directly from the resistance tensor \cite{Samin2018}, and just like $\textsf{\textit{R}}_\mathrm{p}$, they are purely geometrically determined. Furthermore, Eq. \ref{eq:3} captures the coupling between rotation and translation, because the particle path depends on both timescales. The generality of these equations of motion points to their validity for a wide range of particle shapes provided they have at least one plane of mirror symmetry. The equations also hold for particles that do not rotate \--- shapes with more than one mirror plane have an infinitely large $\tau$ and translate at a constant lateral velocity (Supplementary Text 1C).
	
	To test the validity of these equations, we produce a variety of disk dimers and track their motion as they rotate from $\theta\sim0.85\pi$ to $\theta\sim0.10\pi$. Upon comparing the obtained raw experimental trajectories, we see a qualitative similarity (Fig. S4). However, as some particles rotate more slowly than others, the overall paths the particles follow differ considerably in quantitative terms. We fit Eqs. \ref{eq:2} and \ref{eq:3} to the observed trajectories and extract the two characteristic times for each particle, as discussed in Supplementary Text 2. Finally, we transform experimental time to $\left(t-t_\perp\right)/\tau$ for each shape and compare the angle evolution for the set of dimers (top curve in Fig. \ref{fig:3} (a)). The re-orientation dynamics of the studied disk dimers do not only agree quantitatively \--- they seem to be independent of the exact particle shape as evident from the collapsed experimental data, which closely follows Eq. \ref{eq:2}, as well as 3D finite element computations. This apparent shape-independence implies that the characteristic time captures all geometric details of a particle. By condensing them in $\tau$ and factoring them out, we are left with the general dynamics determined by the mirror symmetry and described well by our equation for $\theta\left(t\right)$. This notion is reaffirmed once we take a look at the lateral motion of the disk dimers  (bottom curve in Fig. \ref{fig:3} (a)). Their cross-streamwise motion also appears shape-independent once we use $\left(t-t_\perp\right)/\tau$ instead of experimental time and scale their lateral displacement by the channel height and the characteristic times. Even when the lateral motion of a particle deviates from the one predicted by Eq. \ref{eq:3}, the deviation can be traced back to the re-orientation dynamics. Some dimers stop rotating before their mirror axes align with the flow direction, leading to a decoupling of rotation and translation. Thus, they begin to behave as rods with a finite cross stream velocity even at long timescales \cite{Nagel2018}. A possible reason for these deviations is interaction with hydrodynamic images if the particle comes too close to the wall. Additionally, artefacts of the lithography process such as slight asymmetry in the particle itself or dust of size comparable to $h_\mathrm{g}$, are other possible culprits. We test these notions by simulating the full trajectory of a dimer whose experimental behaviour deviates from the theoretically predicted. Since the 3D finite element results are well-described by the equations of motion and agree with the experimental trajectories, we conclude that the observed deviations are indeed experimental artefacts.
	
	Encouraged by the close agreement between theory and experiments in Fig. \ref{fig:3} (a), we broaden our scope to mirror symmetric particles of various shape. Substituting the disks with pointy building blocks such as squares and triangles leads to different timescales, but does not affect the general particle dynamics (Fig. \ref{fig:3} (b)). Increasing the number of building blocks has the same effect \--- trimers with different size ratios and inter-disk angle also behave identically once we isolate the geometrical details condensed in $\tau$ and $\tau_y$. This universality, remarkable as it is, is not entirely unexpected \--- Eqs. \ref{eq:2} and \ref{eq:3} are derived with the sole assumptions of a force- and torque-free particle with a mirror plane moving in creeping flow. Moreover, our findings suggest we should expect this type of dynamics from any particle that has at least one mirror plane and is subjected to confined Stokes flow. Our reasoning also raises the question what is the behaviour of an asymmetric particle, for instance, a trimer where all three disks have different radii (Fig. S3). One such shape rotates until it acquires a stable orientation $\theta_\infty\neq0$ as discussed in Supplementary Text 1C. However, since the flow disturbance it creates is asymmetric, the particle has a non-zero lateral velocity even after it has ceased re-orienting \cite{Samin2018,Bechert2019}.
	
	\begin{figure*}[ht!]
		\includegraphics[width=\linewidth]{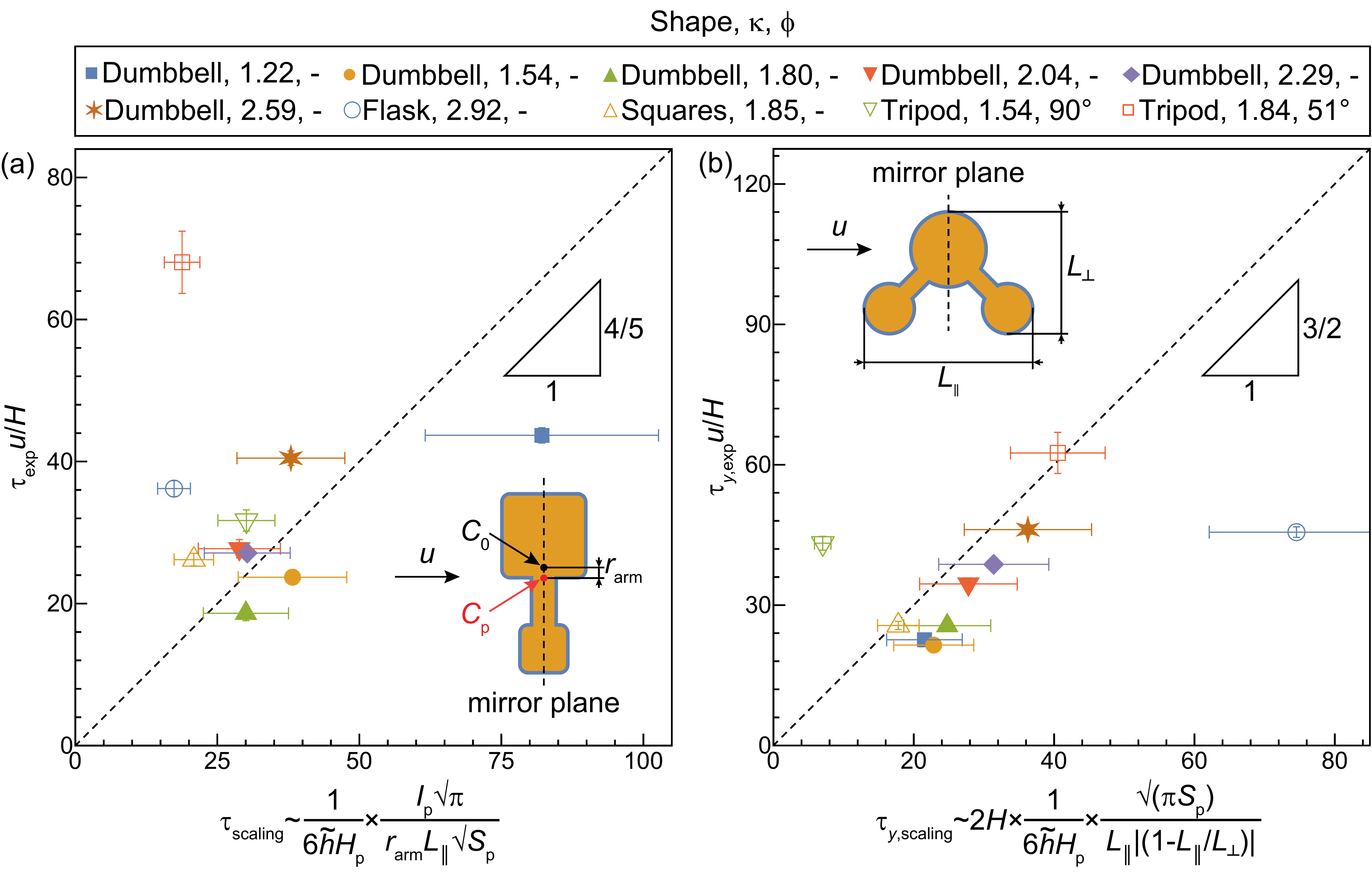}
		\caption{\textbf{Relation of the characteristic timescales to particle geometry.} The (a) rotation and (b) translation timescales needed to fully describe particle motion via Eqs. \ref{eq:2} and \ref{eq:3} are solely dependent on the geometry of the system. For identical flow parameters such as depth-averaged flow velocity $u$, gap thickness $h_\textrm{g}$ and channel height $H$, the detailed shape of the particle determines $\tau$ and $\tau_y$. The rotational timescale depends on the polar moment of inertia $I_\mathrm{p}$ of the particle, its area $S_\mathrm{p}$ (yellow particle sketch), its projected length when perpendicular to the flow $L_\perp$ and the distance $r_\mathrm{arm}$ spanning  from the centroid $\textit{C}_0$ to the centre of perimeter $\textit{C}_\textrm{p}$. We obtain the translational timescale via the area of the particle and its projected lengths $L_\perp$ and $L_\parallel$ when its mirror plane is perpendicular or parallel to the flow, respectively. The vertical error bars represent the standard deviation of the timescales within an experimental series (Table S1). The horizontal error bars are calculated from the uncertainty of the confinement $\tilde{h}$. The dashed lines are a guide to the eye.}\label{fig:4}
	\end{figure*}
	
	Though we have a rigorous description of the general trajectory of a mirror-symmetric particle, its exact motion still depends on two timescales. Up to now we obtain $\tau$ and $\tau_y$ as fitting parameters in Eqs. \ref{eq:2} and \ref{eq:3}. However, knowing their values \textit{a priori} opens the door towards tailoring the shape of a particle to a desired trajectory. One possible way to obtain this target-specific shape is to survey a large variety of particles, compute their resistance tensors and estimate $\tau$ and $\tau_y$ \cite{Samin2018}. As robust as this method is, it is not particularly insightful as it does not yield an explicit relation between the timescales and a particle's geometric parameters. By considering imbalanced rods, we propose scaling arguments linking the timescales $\tau$ and $\tau_y$ of a particle to its geometry. 
	
	We do so by first identifying $\tau$ and $\tau_y$ are functions of the particle velocities $\dot{\theta}$ and $\dot{x}$ at specific orientations: $\tau_\mathrm{scaling}=-1/\dot{\theta}\left(\theta=\pi/2\right)$ and  $\tau_{y\mathrm{,scaling}}=2H/(\dot{x}_\perp-\dot{x}_\parallel)$ as discussed in Supplementary Text 1D. The subscripts of the streamwise velocities denote particle orientation: $\dot{x}_\perp=\dot{x}\left(\theta=\pi/2\right)$ and $\dot{x}_\parallel=\dot{x}\left(\theta=0\right)$. To compute the three velocities, we make use of the force- and torque-free nature of the particle. At any instant in time, the angular momentum it gains from the in-plane flow is dissipated as Couette torque from the confining walls above and below its faces: $T_\mathrm{f}+T_\mathrm{w}=0$. We write a similar balance for the streamwise force \--- the drag from the surrounding fluid and the friction from the confining walls cancel: $F_{x\mathrm{,f}}+F_{x,\mathrm{w}}=0$. In Supplementary Text 4 we propose linear scaling expressions for each torque and force. We solve the two balances for the three velocities and substitute the solutions in the expressions for the two timescales:  
	\begin{equation}\label{eq:4}
	\tau_\mathrm{scaling}\simeq\frac{1}{6\tilde{h}H_\mathrm{p}}\times\frac{H}{u}\times\frac{\sqrt{\pi}I_\mathrm{p}}{r_\mathrm{arm}L_\parallel\sqrt{S_\mathrm{p}}}
	\end{equation}
	and
	\begin{equation}\label{eq:5}
	\tau_{y,\mathrm{scaling}}\simeq 2H\times\frac{1}{6\tilde{h}H_\mathrm{p}}\times\frac{H}{u}\times\frac{\sqrt{\pi S_\mathrm{p}}}{L_\perp-L_\parallel}\frac{L_\perp}{L_\parallel}.
	\end{equation}
	The proposed scaling relations provide estimates for $\tau$ and $\tau_y$ by simplifying the particle geometry to projections of shape $L_\perp$, $L_\parallel$, moment of inertia $I_\mathrm{p}$, area $S_\mathrm{p}$, as well as other geometrical parameters. These parameters are illustrated in the insets of Fig. \ref{fig:4} and detailed in Table 1.
		\begin{table}[h!]
			\caption{\label{tab:table1}%
				\textbf{Scaling expressions for the longitudinal forces and in-plane torques acting on a particle in confined Stokes flow.} The particle moves at velocity $\dot{x}_i$ while rotating with frequency $\dot{\theta}$ in a fluid with depth-averaged flow velocity $u$. The subscript $i\equiv\perp\vee\parallel$ denotes orientation. The forces and torques depend on the particle geometry through its area $S_\mathrm{p}$, polar moment of inertia $I_\mathrm{p}$, thickness $H_\mathrm{p}$, confinement $\tilde{h}$ and projected length $L_i$. The gap height $h_\mathrm{g}=\tilde{h}H$ is made dimensionless with the height of the channel $H$. The two dimensional projected lengths $L_\perp$ and $L_\parallel$ are sketched in Fig. \ref{fig:1} (g) and Fig. \ref{fig:4} (b). We define $r_\mathrm{arm}$ as the distance between the centroid of a particle $C_0$ and its centre of perimeter $C_\mathrm{p}$ and sketch it in Fig. \ref{fig:4} (a).
			}
		\begin{center}
			\begin{tabular}{rllll}
				&& \textrm{Fluid} &&\textrm{Wall}\\
				\midrule
				$F_i\sim$ && $\displaystyle 12\eta \frac{u}{H^2} H_\mathrm{p}L_\parallel\sqrt{\frac{S}{\pi}}\times\frac{L_i}{L_\perp}$ && $\displaystyle-\frac{2\eta}{h}\dot{x}_iS_\mathrm{p}$ \\[15pt]\\[-5pt]
				$T\sim$ &&  $\displaystyle12\eta \frac{u}{H^2} H_\mathrm{p}L_\parallel\sqrt{\frac{S}{\pi}}\times r_\mathrm{arm}$ && $\displaystyle-\frac{2\eta}{h}\dot{\theta}I_\mathrm{p}$ \\[5pt]
				\bottomrule
			\end{tabular}
			\end{center}
		\end{table}

	We verify the scaling models by comparing our experimental timescales to the ones computed via Eqs. \ref{eq:4} and \ref{eq:5} in Fig. \ref{fig:4}. We complement this comparison with numerical timescales, computed via 3D FEM, and present them in Fig. S18 and S19. The scaling relation for $\tau_\mathrm{scaling}$ overestimates $\tau_\mathrm{exp}$ by a factor of 1.25, while $\tau_{y\mathrm{,scaling}}$ underestimates $\tau_{y\textrm{, exp}}$ by a factor 1.5. This mismatch is to be expected as the proposed minimalistic scalings strip the particles of any geometric detail. A possible remedy is the incorporation of mean particle curvatures, which, however, comes at the expense of model simplicity. 
	
	Though we determine the two timescales up to a scaling factor of order 1, Eqs. \ref{eq:4} and \ref{eq:5} accurately predict when $\tau$ or $\tau_y$ diverge and when $\tau_y$ becomes negative. In some trivial cases, particles cease to rotate and $\tau\to\infty$ when they are either too thick ($\tilde{h}\to0$), too thin ($H_\mathrm{p}\to 0$) or there is no flow ($u\to 0$). The timescale also diverges when the distance between the centroid and the centre of perimeter vanishes ($r_\mathrm{arm}\to 0$). Particles with more than one mirror plane \--- rods, symmetric dimers and disks \--- all have $r_\mathrm{arm}=0$. Similarly, particles do not cross streamlines when their two projected lengths match $L_\perp=L_\parallel$. One such particle is a trimer with $\kappa=1.5$ and $\phi\sim\SI{68}{\degree}$, which rotates without drifting away from the centre-line of the channel, as demonstrated by finite element computations in Fig. \ref{fig:1} (e). We also observe this phenomenon experimentally: the trimer with $\kappa=1.84$ and $\phi=\SI{51}{\degree}$ barely moves in the lateral direction (Fig. \ref{fig:2} (e)). Its large $\tau_y$, dampening its lateral motion, is due to its comparable projected lengths. Furthermore, $\tau_y$ may become negative for trimers with large $\phi$, as demonstrated in Fig. \ref{fig:1} (e). This change in drift direction is present experimentally for a trimer with $\kappa=1.5$ and $\phi\sim\SI{90}{\degree}$ and is the reason why we compare  $\left|\tau_{y\mathrm{, exp}}\right|$ to  $\left|\tau_{y\mathrm{, model}}\right|$ in Fig. \ref{fig:4} (b). 
	
	The applicability of the proposed scaling relations to a wide range of particles with different geometry and symmetry supports the main conclusion of our work: in confined Stokes flow, particles with at least one mirror plane behave identically as long as we scale their trajectories by characteristic times, directly related to their shape. The proposed scaling can be utilized to predict trajectories of particles based on minimalistic scaling arguments.
	
	\section*{Conclusion}
	In summary, by combining experiments, simulations and theory, we investigate how the trajectory of a confined particle subjected to Stokes flow is determined by its geometry. We observe that particles with a single mirror plane exhibit qualitatively similar behaviour: they rotate in-plane to align their mirror axis with the flow and their larger building block upstream, all the while crossing streamlines. However, the timescales over which this dynamics happens are strongly dependent on particle shape. We fit our experimental trajectories and finite element calculations to theoretical equations of motion we have recently derived, thus extracting characteristic rotational and translational times for each particle. By scaling experimental time by the respective rotational timescale for each experiment, we collapse the evolution of the orientation for all particles onto a single curve. Similarly, we obtain a universal bell-shaped path by scaling real time and a particle's cross-streamline velocity. Finally, we propose minimalistic scaling relations linking the characteristic times of a particle to its geometry. We strip the particles of all geometrical details and treat them as imbalanced rods, thus reinforcing the idea that it is solely their symmetry that defines their overall dynamics.  Our observations suggest the trajectories are universal for particles with at least one mirror plane. This finding deepens our understanding of fluid-structure interactions in confined Stokes flow. Moreover, it opens new opportunities in lab-on chip and industrial applications enabling shape-based separation of suspended particles solely through hydrodynamic interactions.
	
	\section*{Methods}
	\subsection*{Experimental setup}
	Polymeric microparticles are produced and observed with an experimental setup, similar to the one used by Uspal, Eral and Doyle \cite{Uspal2013}. Polydimethylsiloxane (PDMS, Sylgard\textregistered 184, Dow Corning) microfluidic devices of width $W=512\pm\SI{2}{\micro\metre}$ are fabricated according to Dendukuri \textit{et al.} \cite{dendukuri2008pdms}. Disk dimers are tracked in channels with height $H=30\pm\SI{1}{\micro\metre}$. Trimers, triangle and square dimers are tracked in a 33-micron high channels. A UV-crosslinking oligomer, poly-(ethyleneglycol) diacrylate (PEG-DA   $M_n=\SI{700}{}$ ,  $\eta=\SI{95}{\milli\pascal\second}$ , Sigma-Aldrich), is mixed with a photoinitiator, hydroxy-2-methylpropiophenone, (Darocur\textregistered 1173, Sigma-Aldrich), in a 19:1 volume ratio and the mixture is pumped through the microfluidic channel. The device, loaded with prepolymer, is mounted on the stage of a motorized Nikon Ti Eclipse inverted optical microscope. A photolithographic mask with well-defined shape is inserted as a field stop. Mask designs are made in Wolfram Mathematica\textregistered and post-processed in Dassault Systémes’ DraftSight\textregistered.
	
	\subsection*{Particle production and tracking}
	
	Microparticles are produced by shining a \SI{100}{\milli\second} pulse of UV light through the mask onto the channel, thus confining photopolymerization to a discrete part of the prepolymer mixture. Oxygen, diffusing through the permeable PDMS walls of the device, inhibits polymerization in their vicinity \cite{dendukuri2008pdms}. This facilitates the formation of two thin lubrication layers, $h_\mathrm{g}=2.5\pm\SI{0.5}{\micro\metre}$, which separate the particles from the confining walls of the channel. Particles are produced and observed with a 20X lens. The microparticle is set in motion by applying a pressure drop  $\Delta p\approx\SI{1.5}{\kilo\pascal}$  across the channel resulting in a depth-average flow velocity $u=\SI{55}{\micro\metre\per\second}$ for the shallower channel and $u=\SI{70}{\micro\metre\per\second}$ for the 33-micron high channel. The particle is tracked by moving the automated microscope stage in a stepwise manner.
	
	The positions and orientations of particles containing disks are extracted from the acquired time series using a custom-written MATLAB script, which employs circular Hough transforms to identify the particle shape in each frame. The script utilizes MATLAB’s Bio-Formats package \cite{Linkert2010} and the calcCircle tool. Particles comprising triangles and squares are tracked by fitting an ellipse to them, calculating the angle and detecting their straight edges.
	
	\subsection*{Finite element computations}
	All computational results are obtained through the Finite element method as implement in the Creeping Flow module of COMSOL Multiphysics 5.3, which we couple to MATLAB via LiveLink. Each solution is carried out on a single computational node fitted with an Intel Xeon E5-2620 v4 @ 2.10GHz CPU and 64 GB memory. Technical details regarding geometry building, meshing and solver settings are given in Supplementary Text 3 \cite{MUMPS1,MUMPS2,Holzbecher2008}. 
	
	We use the channel height $H=1$ as a length scale. We set the inlet flow velocity $u$, the kinematic viscosity of the fluid $\eta$ and its mass density $\rho$ to unity. To simulate creeping flow at this $Re=1$, we neglect the inertial term in the momentum equation and solve the Stokes equation with no external forcing:
	
	\begin{equation*}
		\nabla\cdot\left(-p\textsf{\textit{I}}+\eta\left(\nabla\bm{U}_\mathrm{f}+\nabla\bm{U}_\mathrm{f}^\intercal\right)\right)=0
	\end{equation*} 
	\begin{equation*}
		\nabla\cdot\bm{U}_\mathrm{f}=0,
	\end{equation*}
	where we solve for $\bm{U}_\mathrm{f}$ and $p$, the fluid velocity and pressure fields. We integrate the total stress over the particle surface to obtain the forces and torque acting on it at a given position and orientation with respect to the flow. To compute the force- and torque-free velocities of the particle at this configuration, we numerically solve the force balance:
	
	\begin{equation*}
		\begin{pmatrix}
			\dot{x}\\
			\dot{y}\\
			\dot{\theta}
		\end{pmatrix}=-\frac{1}{\mu}\textsf{\textit{R}}_\mathrm{p}^{-1}\cdot\bm{F}_0,
	\end{equation*}
	where $\bm{F}_0$ is the forces and torque acting on a stationary particle in a flow and $\textsf{\textit{R}}_\mathrm{p}$ is the resistance tensor for this configuration (Supplementary Note 1A, equation (1)). We obtain the trajectory of a particle through a first order time integration scheme, where we apply $\left(\dot{x},\dot{y},\dot{\theta}\right)$ over a timestep $t_\text{step}$, which we determine every iteration (Supplementary Text 3).




\end{document}